\documentclass{article}
\usepackage{spconf,amsmath,graphicx}
\usepackage[hidelinks]{hyperref}
\usepackage[sort]{cite}
\usepackage{tikz}
\usetikzlibrary{arrows.meta,calc}
\usepackage{booktabs} 
\usepackage{adjustbox} 
\usepackage{amssymb}  
\usepackage{makecell}
\usepackage{diagbox}
\usepackage{siunitx}
\usepackage{multirow} 
\usepackage{enumitem}
\usepackage{bm}  
\usepackage{pbalance}

\usepackage{acronym}
\acrodef{DNN}{deep neural network}
\acrodef{GAN}{generative adversarial network}
\acrodef{DoA}{direction of arrival}
\acrodef{PDF}{probability density function}
\acrodef{E2E}{end-to-end}
\acrodef{STFT}{short-time Fourier transform }
\acrodef{TF}{time-frequency}
\acrodef{RIR}{room impulse response }
\acrodef{FiLM}{feature-wise linear modulation }
\acrodef{SNR}{signal-to-noise ratio}
\acrodef{MOS}{mean opinion score}
\acrodef{DNN}{deep neural network}
\acrodef{PDF}{probability density function}
\acrodef{FiLM}{feature-wise linear modulation }
\acrodef{TSE}{Target speaker extraction }
\acrodef{SE}{speech enhancement}

\newcommand{\Xv}{\mathbf{X}}
\newcommand{\xv}{\mathbf{x}}
\newcommand{\sv}{\mathbf{s}}
\newcommand{\hatsv}{\hat{\mathbf{s}}}
\newcommand{\Hv}{\mathbf{H}}
\newcommand{\Dl}{\mathbf{D}_{\text{L}}}
\newcommand{\doa}{\varphi}
\newcommand{\varphiParam}{\Theta}

\newcommand{\SpatialGAN}{\text{SpatialGAN}}

\newcommand{\Stftxv}{\mathbf{X}_\text{F}}
\newcommand{\Gl}{\mathbf{G}_\text{L}}

\newcommand{\RealMat}[2]{\mathbb{R}^{#1 \times #2}}
\newcommand{\LD}{\mathcal{L}_\text{D}}
\newcommand{\LG}{\mathcal{L}_\text{G}}
\newcommand{\hatStft}{\widehat{\mathbf{S}}} 

\newcommand{\RealTensor}[3]{\mathbb{R}^{#1 \times #2 \times #3}} 

\newcommand{\Lrec}{\mathcal{L}_{\text{rec}}}
\newcommand{\Ladv}{\mathcal{L}_{\text{adv}}}
\newcommand{\Lfeat}{\mathcal{L}_{\text{feat}}}
\newcommand{\Lt}{\mathcal{L}_\text{t}}
\newcommand{\Lf}{\mathcal{L}_\text{f}}

\def\gen{\mathcal{G}}
\newcommand{\disc}{\mathcal{D}}
\newcommand{\concat}[1]{\mathrm{concat}\left( #1 \right)}
\newcommand{\DecFeat}{\mathbf{D_\text{f}}} 
\newcommand{\EncFeat}{\mathbf{E_\text{f}}} 
\newcommand{\AttFeat}{\mathbf{A_\text{f}}} 
\newcommand{\tildedl}{\widetilde{\mathbf{D}}_{\text{L}}}

\newcommand{\gammar}{\bm{\gamma}}
\newcommand{\betar}{\bm{\beta}}

\title{GAN-Based Multi-Microphone Spatial Target Speaker Extraction}
\name{Shrishti Saha Shetu$^{1}$, Emanu\"{e}l A. P. Habets$^{1}$, Andreas Brendel$^2$}

\address{
$^{1}$International Audio Laboratories Erlangen\footnotemark[1], Am Wolfsmantel 33, 91058 Erlangen, Germany\\
$^{2}$Fraunhofer IIS, Am Wolfsmantel 33, 91058 Erlangen, Germany\\
{\small \{shrishti.saha.shetu, emanuel.habets, andreas.brendel\}@iis.fraunhofer.de}
}

\begin{document}
\ninept
\maketitle

\footnotetext[1]{A joint institution of Fraunhofer IIS and Friedrich-Alexander-Universit{\"a}t Erlangen-N{\"u}rnberg (FAU), Germany.}

\begin{abstract}
Spatial target speaker extraction isolates a desired speaker’s voice in multi-speaker environments using spatial information, such as the \ac{DoA}. Although recent \ac{DNN}-based discriminative methods have shown significant performance improvements, the potential of generative approaches, such as generative adversarial networks (GANs), remains largely unexplored for this problem. In this work, we demonstrate that a GAN can effectively leverage both noisy mixtures and spatial information to extract and generate the target speaker’s speech. By conditioning the GAN on intermediate features of a discriminative spatial filtering model in addition to \ac{DoA}, we enable steerable target extraction with high spatial resolution of 5$^\circ$, outperforming state-of-the-art discriminative methods in perceptual quality–based objective metrics.
\end{abstract}
\section{Introduction}
\label{sec:intro}

Humans can selectively attend to a single sound source in complex acoustic scenes, effectively suppressing competing speakers and background noise, a phenomenon known as the cocktail party effect \cite{cherry1953some, pollack1957cocktail}. \ac{TSE} replicates this selective attention by isolating a target speaker’s voice from noisy and reverberant mixtures, which is critical for applications such as hearing aids, conference systems, and automatic speech recognition \cite{kovalyov2023dsenet, 10694623, li2014overview}. To achieve this, researchers leverage spatial information \cite{5109760,gannot2017consolidated,5067369,gu2019neural,benesty2008microphone}, visual features \cite{ephrat2018looking,alfouras2018conversation,shetu2021empirical}, and speaker embeddings \cite{zmolikova2023neural,elminshawi2024new,ji2020speaker} mainly to identify the target speaker in the observed mixture.

\begin{figure}[t]
\centering
\begin{tabular}{@{}c@{\hspace{0.05\linewidth}}c@{}}
\begin{minipage}{0.55\linewidth}
\centering
\begin{tikzpicture}[scale=1.45]
\def\rin{0.7}      
\def\rout{1.2}     
\def\rmid{0.95}    
\def\phit{50}        
\def\targetwidth{10} 
\def\rarray{0.15}    
\def\lext{1.6}       

\fill[gray!15,even odd rule] (0,0) circle (\rout) (0,0) circle (\rin);

\fill[white] 
  ({\rin*cos(\phit-\targetwidth)},{\rin*sin(\phit-\targetwidth)}) 
  arc[start angle=\phit-\targetwidth,end angle=\phit+\targetwidth,radius=\rin] --
  ({\rout*cos(\phit+\targetwidth)},{\rout*sin(\phit+\targetwidth)}) 
  arc[start angle=\phit+\targetwidth,end angle=\phit-\targetwidth,radius=\rout] -- cycle;

\draw[dotted,thick] (0,0) circle (\rin);
\draw[dotted,thick] (0,0) circle (\rout);

\pgfmathsetmacro{\graystart}{\phit+\targetwidth} 
\pgfmathsetmacro{\grayend}{\phit+\targetwidth+340} 
\pgfmathsetmacro{\sectorangle}{340/5} 

\foreach \i in {0,1,2,3,4} {
    \pgfmathsetmacro\sa{mod(\graystart + \i*\sectorangle,360)}
    \draw[black,thick] (\sa:\rin) -- (\sa:\rout);
}

\draw[gray] (0,0) circle (\rarray);
\foreach \ang in {30,160,290}{
  \fill ({\rarray*cos(\ang)},{\rarray*sin(\ang)}) circle (1.8pt);
}

\foreach \i in {0,1,2,3,4} {
    \pgfmathsetmacro\ang{mod(\graystart + (\i+0.5)*\sectorangle,360)} 
    \begin{scope}[shift={({\rmid*cos(\ang)},{\rmid*sin(\ang)})},scale=0.12]
        \draw[fill=black] (0,0.6) circle (0.3);
        \draw[line width=1pt,line cap=round] (0,0.3) -- (0,-0.6);
        \draw[line width=1pt,line cap=round] (-0.3,0) -- (0.3,0);
        \draw[line width=1pt,line cap=round] (0,-0.6) -- (-0.2,-1.0);
        \draw[line width=1pt,line cap=round] (0,-0.6) -- (0.2,-1.0);
    \end{scope}
}

\begin{scope}[shift={({\rmid*cos(\phit)},{\rmid*sin(\phit)})},scale=0.12]
    \draw[fill=red] (0,0.6) circle (0.3);
    \draw[red,line width=1.5pt,line cap=round] (0,0.3) -- (0,-0.6);
    \draw[red,line width=1.5pt,line cap=round] (-0.3,0) -- (0.3,0);
    \draw[red,line width=1.5pt,line cap=round] (0,-0.6) -- (-0.2,-1.0);
    \draw[red,line width=1.5pt,line cap=round] (0,-0.6) -- (0.2,-1.0);
\end{scope}

\draw[dashed,gray] (0,0) -- (\lext,0);
\draw[dashed,gray] (0,0) -- ({\lext*cos(32)},{\lext*sin(32)});
\draw[dashed,blue,thick] (0,0) -- ({\lext*cos(\phit)},{\lext*sin(\phit)});

\pgfmathsetmacro{\midphim}{0.5*32}
\pgfmathsetmacro{\midphit}{0.5*(32+\phit)}
\pgfmathsetmacro{\Rarc}{\rout+0.25} 

\draw[->,gray] ({\Rarc},0) arc[start angle=0,end angle=32,radius={\Rarc}];
\node[gray] at ({(\Rarc+0.15)*cos(\midphim)},{(\Rarc+0.15)*sin(\midphim)}) {$\varphi_m$};

\draw[->,blue] ({\Rarc*cos(32)},{\Rarc*sin(32)}) arc[start angle=32,end angle=\phit,radius={\Rarc}];
\node[blue] at ({(\Rarc+0.15)*cos(\midphit)},{(\Rarc+0.15)*sin(\midphit)}) {$\doa$};

\draw (-1.6,-1.25) rectangle (1.8,1.45);

\end{tikzpicture}
\end{minipage}
&
\hspace{.2cm}
\begin{minipage}{0.38\linewidth}
\footnotesize
\begin{tabular}{@{}ll@{}}
\toprule
\multicolumn{2}{l}{\textbf{Room characteristics}} \\
\midrule
Width & 2.5--5 m \\
Length & 3--9 m \\
Height & 2.2--3.5 m \\
T60 & 0.2--0.5 s \\
\bottomrule
\end{tabular}
\end{minipage}
\end{tabular}

\caption{
Simulation setup showing the target source at a random angle $\doa$ relative to the microphone orientation $\varphi_m$, with five interfering speakers placed in separate gray areas.
}
\label{fig:Simulation}
\label{fig:simulation_setup}
\end{figure}

Many recent multi-microphone \ac{DNN}-based \ac{TSE} approaches exploit spatial guidance. They either use adaptive beamforming to extract target speaker features for single-channel \ac{DNN}-based \ac{TSE} \cite{elminshawi2023beamformer} or train multichannel \ac{DNN}s \ac{E2E} with spatial cues, such as the \ac{DoA}, as conditioning information \cite{TeschJNF,tesch2023multi,jing2025end}. However, all spatially guided \ac{DNN}-based methods,  to the best of our knowledge, rely on discriminative training \cite{jebara2012machine}. Although the discriminative methods show signal improvements in different \ac{SE} tasks, including \ac{TSE}, their performance in low \ac{SNR} environments remains limited \cite{shetu2024comparative}. In contrast, generative  approaches promise the potential
for superior performance in these scenarios by learning the distribution of clean speech signals.

Recently, generative approaches, particularly \ac{GAN}-based methods, have proven effective and often outperform discriminative models in speech enhancement, synthesis, and restoration tasks \cite{fu2019metricgan, sani2023improving,shetu2025gan,chen2025tfdense,yu2024ks}. While speaker embedding–based generative \ac{TSE} has been explored \cite{10446418,kamo2023target,ma2025enhancing,navon2025flowtse}, to our knowledge, no prior work has investigated whether a generative model can leverage spatial information to extract a target speaker and generate clean speech. In this work, we condition a \ac{GAN} on spatial information, such as the \ac{DoA}, to learn the conditional distribution of the target speaker’s clean speech. Our specific contributions are:

\begin{itemize}[noitemsep, leftmargin=*]
    \item We demonstrate that a \ac{GAN} can effectively leverage spatial information and be trained \ac{E2E} for both fixed-target scenarios, with a fixed geometric setting, and steerable scenarios, where it can be steered toward any desired direction \cite{TeschJNF}, achieving better performance to state-of-the-art discriminative spatial \ac{TSE} methods.

    \item We investigate different conditioning approaches for \ac{GAN}-based spatial \ac{TSE}, and show that incorporating intermediate features from a discriminative spatial filtering model together with \ac{DoA} information provides more effective conditioning and better performance  than relying on \ac{DoA} alone.
\end{itemize}

\begin{figure*}[t]
\centering
\includegraphics[width=.86\textwidth]{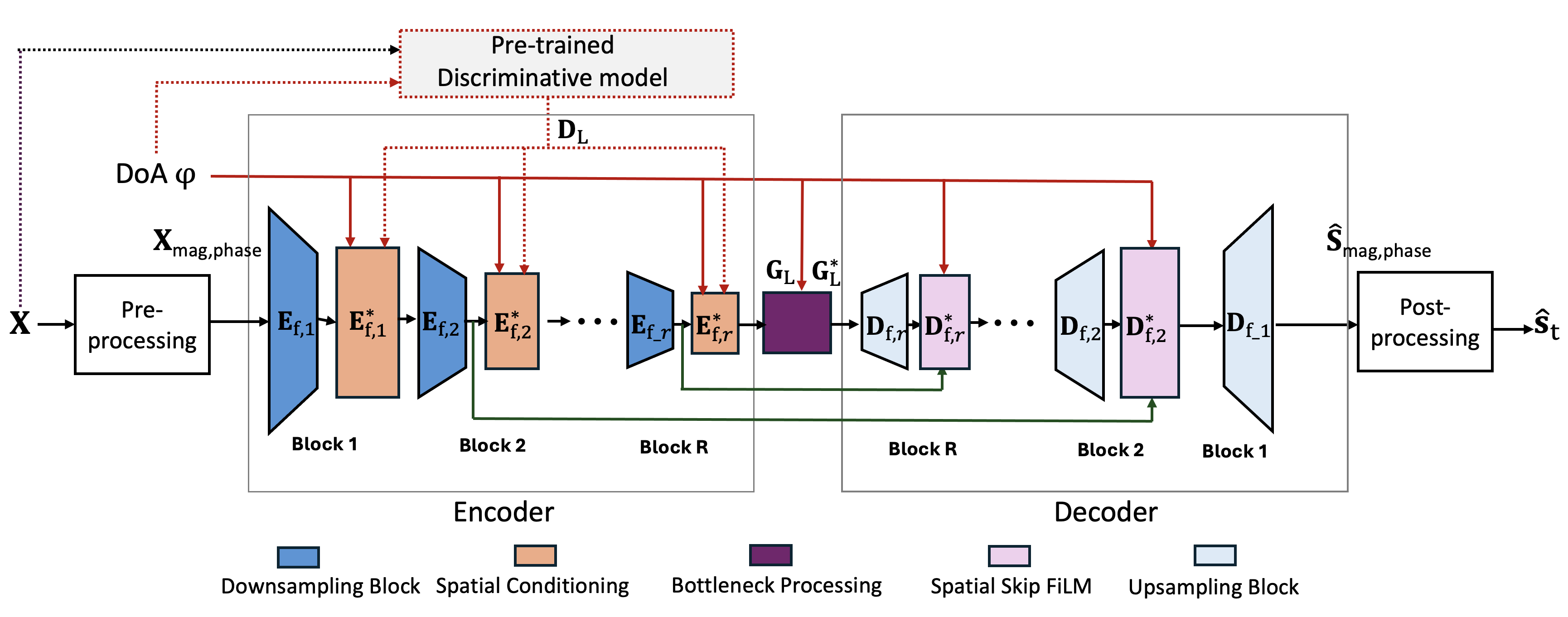}
\caption{ Overview of  the  proposed spatially conditioned \ac{GAN}  ($\SpatialGAN$).}
\label{fig:LF}
\end{figure*}

\section{Problem Formulation}
\label{sec:PF}
We consider a multi-speaker scenario where a circular array of $M$~omnidirectional microphones records a target speaker at direction $\doa$ along with $K$ interferers at distinct spatial positions, as illustrated in Fig.~\ref{fig:simulation_setup}. Let $\sv_\text{t} \in \mathbb{R}^L$ denote the discrete-time dry target signal, where \( L \) denotes the length of the signal in samples, $\sv_k \in \mathbb{R}^L$ the $k$-th interferer, and $\Hv_{m,\textrm{t}}$, $\Hv_{m,k}$ the room impulse responses (RIRs) from the target and the $k$-th interferer to microphone $m$. Then, the signal observed at the $m$-th microphone is given by,
%
\begin{equation}
\xv_m = \Hv_{m,\textrm{t}} * \sv_\text{t} + \sum_{k=1}^{K} \Hv_{m,k} * \sv_k, 
\quad m = 1, \dots, M,
\end{equation}
where $*$ denotes discrete-time convolution. In the following, all interferers and reverberation are treated as noise. The objective is to recover the dry target signal $\sv_\text{t}$ from the reverberant multi-channel recordings $\Xv = [\xv_1, \dots, \xv_M]^\top \in \mathbb{R}^{M \times L}$.

\section{Proposed Method}
\label{sec:PM}

\subsection{GAN Formulation}
\label{sec:GF}
Assuming a discrete spatial resolution for the \ac{DoA} $\doa$, conventional \ac{E2E} discriminative \ac{DNN}-based approaches extract the target signal as $ \sv_\text{t} = \mathcal{F}(\Xv, \doa)$, where $\mathcal{F}$ denotes the discriminative \ac{DNN} model. In contrast, our work investigates an \ac{E2E} spatially-conditioned \ac{GAN} method, referred to as $\SpatialGAN$. The generator models the conditional distribution of the target speech given the mixture and the spatial information as, \begin{equation} 
\sv_\text{t} \sim p_{\text{G}}(\sv_\text{t} \mid \Xv, \doa),
\end{equation}
where $p_{\text{G}}$ is learned to approximate the true conditional distribution of the target signal. We further explore conditioning on intermediate discriminative features $\Dl$ extracted from $\mathcal{F}$ inspired by \cite{shetu2025leveraging}, or jointly on both $\doa$ and $\Dl$, in which case the generator models  
\begin{equation}
\sv_\text{t} \sim p_{\text{G}}(\sv_\text{t}\mid \Xv, \doa, \Dl),
\end{equation}
depending on the available conditioning information. We hypothesize that intermediate features obtained from a discriminative spatial filtering model provide complementary \ac{TF} representations of the target speaker, which can be beneficial for the $\SpatialGAN$.  The learning objective is formulated as
\begin{equation}
\min_{\mathbb{\theta}} \max_{\varphiParam} \; 
\mathbb{E}_{\sv_\text{t} } [ \LD(\varphiParam, \sv_\text{t}) ] 
+ \mathbb{E}_{\Xv} [ \LG(\varphiParam, \mathbb{\theta}, \Xv, \doa, \Dl) ],
\end{equation}
with $\LD$ and $\LG$ as the discriminator and generator losses, and $\mathbb{\theta}$ and $\varphiParam$ the generator and discriminator parameters.

\subsection{Loss Functions}
\label{sec:LO}
The training objective combines reconstruction, adversarial, and feature-matching losses: $\Lrec$, $\Ladv$, and $\Lfeat$. The reconstruction loss consists of a time-domain term $\Lt$, which minimizes an $\ell_1$ loss, and frequency-domain terms $\Lf$, which minimize both $\ell_1$ and Frobenius distances on Mel and magnitude spectra at multiple resolutions \cite{shetu2025leveraging,du2023funcodecfundamentalreproducibleintegrable}. The adversarial loss comprises generator and discriminator components:
\begin{equation}
\resizebox{\columnwidth}{!}{$
\Ladv(\hatsv_\text{t}) = \mathbb{E}_{\Xv, \doa, \Dl} \Big[ 
\frac{1}{K} \sum_{k,n=1}^{K,N_k} \frac{1}{N_k} 
\left[ 1 - \disc_{k,n}(\gen(\Xv,\doa, \Dl)) \right]_+ 
\Big]
$}
\end{equation}

\begin{equation}
\resizebox{\columnwidth}{!}{$
\begin{aligned}
\LD(\sv_\text{t}, \hatsv_\text{t}) ={} & 
\mathbb{E}_{\sv_\text{t}} \Big[ 
\frac{1}{K} \sum_{k,n=1}^{K,N_k} \frac{1}{N_k} 
\left[ 1 - \disc_{k,n}(\sv_\text{t}) \right]_+ \Big] \\
& + \mathbb{E}_{\Xv, \doa, \Dl} \Big[ 
\frac{1}{K} \sum_{k,n=1}^{K,N_k} \frac{1}{N_k} 
\left[ 1 + \disc_{k,n}(\gen(\Xv,\doa, \Dl)) \right]_+ \Big],
\end{aligned}
$}
\end{equation}
where $[\mathbf{x}]_+ = \max(0, \mathbf{x})$, $K$ is the number of discriminators, $N_k$ the number of frames for discriminator $k$, and $\disc_{k,n}$ is the $k$th discriminator at frame $n$. The feature matching loss compares intermediate discriminator features between the target and generated signals:
\begin{equation}
\resizebox{\columnwidth}{!}{$
\Lfeat(\sv_\text{t}, \hatsv_\text{t}) = 
\mathbb{E}_{\sv_\text{t}, \hatsv_\text{t}} \Big[ 
\frac{1}{K L} \sum_{k,n,l=1}^{K,N_k,L} \frac{1}{N_k} 
\big\| \disc^{(l)}_{k,n}(\sv_\text{t}) - \disc^{(l)}_{k,n}(\hatsv_\text{t}) \big\|_1 \Big],
$}
\end{equation} 
with $L$ the number of discriminator layers and $\disc^{(l)}_{k,n}$ the output of layer $l$ at frame $n$ for discriminator $k$. The total generator loss is $
\LG = \Lrec + \lambda_{\text{adv}} \Ladv + \lambda_{\text{feat}} \Lfeat $, where $\lambda_{\text{adv}}$ and $\lambda_{\text{feat}}$ weight the adversarial and feature matching terms.

\section{Model Architecture}
\label{MA}
In our work, we employ a SEANet-based architecture as the generator \cite{tagliasacchi2020seanetmultimodalspeechenhancement}, adopted from \cite{shetu2025gan,shetu2025leveraging}. The architecture follows a U-Net-like structure in the \ac{TF} domain with a symmetric encoder--decoder network and skip-connections. The generator input is derived from the \ac{STFT} of the multi-microphone mixture signal $\Stftxv = \mathrm{STFT}(\Xv)$ by computing the log-magnitude and real and imaginary part of the phase terms $\Xv_{\text{mag},\text{phase}} \in \RealTensor{3M}{F}{T}$, where $F$ denotes the number of frequency bins and $T$ denotes the number of time frames. These representations are concatenated along the first (channel) dimension, i.e.,
\begin{equation}
\begin{aligned}
\Xv_{\textrm{mag,phase}} &= \concat{ \log(|\Stftxv|), \frac{\Re(\Stftxv)}{|\Stftxv|}, \frac{\Im(\Stftxv)}{|\Stftxv|}} .
\end{aligned}
\end{equation}
The generator $\gen$ outputs estimated target speaker's clean speech spectral features $\hatStft_{{\text{mag},\text{phase}}}  \in \RealTensor{3}{F}{T}$, which consist of the magnitude and phase components $\left[ \hatStft_{\textrm{mag}}, \hatStft_{\textrm{r}}, \hatStft_{\textrm{i}} \right]$. The estimated time-domain speech signal $\hatsv_\text{t}$ is obtained as
\begin{equation}
\hatsv_\text{t} = \mathrm{ISTFT} \left( \mathrm{Softplus}(\hatStft_{\textrm{mag}}) \odot \left( \hatStft_{\textrm{r}} + j \hatStft_{\textrm{i}} \right) \right).
\end{equation}

For the discriminator architecture, we adopt a multi-scale STFT-based network  following~\cite{defossez2022high} and with similar parameterization of ~\cite{shetu2025leveraging, du2023funcodecfundamentalreproducibleintegrable}. 
\subsection{Encoder and Decoder}
The encoder consists of eight downsampling $2$D convolutional blocks similar to ~\cite{shetu2025leveraging,du2023funcodecfundamentalreproducibleintegrable} with a maximum number of channels $C=384$. Each downsampling is followed by a spatial conditioning block (Sec.~\ref{SpatialCond}) to fuse the encoder's intermediate representation with the \ac{DoA} and discriminative features $\Dl$. The generator's latent representation $\Gl \in \RealMat{256}{T}$ is further processed in the bottleneck layer with LSTMs and is conditioned further with \ac{DoA}. The decoder mirrors the encoder architecture, featuring \ac{DoA}-conditioned \ac{FiLM} for skip connections (Sec.~\ref{BPandSKF}), followed by an upsampling block comprising 2D transpose convolutions.

\subsection{Spatial Conditioning}
\label{SpatialCond}
In each encoder block $r$, the encoder features $\EncFeat_{,r} \in \RealTensor{C_r}{F_r}{T}$ are conditioned on the discrete \ac{DoA} $\doa$ and the intermediate discriminative features $\Dl \in \RealTensor{C'}{F'}{T'}$. 
The \ac{DoA} is one-hot encoded as $\mathbf{p}_\doa \in \mathbb{R}^D$, where $D$ denotes the total number of discrete spatial positions, and is projected through a linear layer to match the encoder frequency dimension $F_r$ and repeated over time frames $T$, yielding 
$\mathbf{P}_{\doa, r} \in \RealTensor{C_r}{F_r}{T}$. The discriminative features $\Dl$ are linearly interpolated to match the encoder features time dimension $T$ and are subsequently projected via a $2$-D strided convolution to $\tildedl \in \RealTensor{C_r}{F_r}{T}$. We then use an attention layer, where the combined features $\tildedl + \mathbf{P}_{\doa, r}$ serve as queries, while the projected encoder features $\EncFeat'_{,r}$ serve as keys and values, and an estimate of the \ac{TF} feature representation of the target speaker is obtained from the attention output
\begin{equation}
    \AttFeat_{,r} = \mathrm{Softmax}\!\left( \frac{(\tildedl+\mathbf{P}_{\doa, r}) \odot \EncFeat'_{,r}}{\sqrt{C_r}} \right) \odot \EncFeat'_{,r}.
\end{equation}
To further refine the representation, $\AttFeat_{,r}$ is also used to estimate a bounded mask $\mathcal{M}_r \in [0,2]^{C_r \times F_r \times T}$, and the final conditioned features are given by, 
\begin{equation} 
\EncFeat_{,r}^\star = \EncFeat_{,r} \odot \mathcal{M}_r + \AttFeat_{,r}.
 \end{equation}

\subsection{Bottleneck Processing and Spatial Skip FiLM}
\label{BPandSKF}
We also condition the generator latent features $\Gl$ with the \ac{DoA} $\doa$ in the bottleneck processing block using \ac{FiLM}~\cite{perez2018film}. 
The one-hot encoded vector $\mathbf{p}_\doa$ is first embedded and then repeated over time to obtain $
    \mathbf{P}_{\doa\text{L}} \in \RealMat{256}{T}$,
so as to match the dimensionality of $\Gl$, and the conditioned latent representation is computed as,
\begin{equation} 
    \Gl^\star = \Gl + (\gammar_\text{L} \odot \Gl + \betar_\text{L}),
    \end{equation}
where $\gammar_\text{L}$ and $\betar_\text{L}$ denote the FiLM scale and shift parameters, respectively, obtained through $\mathrm{ReLU}$ and $\tanh$ activations.

Following \cite{shetu2025leveraging}, we also condition the decoder features $\DecFeat_{,r} \in \RealTensor{C_r}{F_r}{T_r}$ with the corresponding encoder features $\EncFeat_{,r}$ using \ac{FiLM}. 
To further incorporate \ac{DoA} information for improved target speaker selectivity, we first concatenate the projected one-hot \ac{DoA} embedding $\mathbf{P}_{\doa, r}$ with the encoder features $\EncFeat_{,r}$ along the channel dimension. 
Two separate $2$D convolutional layers then produce the scale and shift factors $\gammar_r$ and $\betar_r$, and the conditioned decoder features are computed as,
\begin{equation} 
\DecFeat_{,r}^\star = \DecFeat_{,r} + \left(\gammar_r \odot \DecFeat_{,r} + \betar_r \right).
\end{equation}

\begin{table}[!t]
\caption{PESQ, SegSNR (dB), and SCOREQ (MOS) results for the fixed-target scenario.}
\centering
\scriptsize
\resizebox{\columnwidth}{!}{%
\begin{tabular}{lcccccc}
\toprule
\textbf{Model} 
& \multicolumn{2}{c}{\textbf{PESQ (↑)}} 
& \multicolumn{2}{c}{\textbf{SegSNR (↑)}} 
& \multicolumn{2}{c}{\textbf{SCOREQ (↑)}} \\
\cmidrule(lr){2-3} \cmidrule(lr){4-5} \cmidrule(lr){6-7}
& -5 dB & 0 dB & -5 dB & 0 dB & -5 dB & 0 dB \\
\midrule
Noisy        & 1.07 & 1.08 & -10.03 & -5.92 & 1.23 & 1.41 \\
JNF \cite{tesch2023multi}         & 1.73 & 2.17 & 6.89   & 9.12  & 2.69 & 3.38 \\
DSENet \cite{jing2025end}      & 1.78 & 2.22 & \textbf{7.70} & \textbf{9.95} & 2.88 & 3.47 \\
SpatialGAN   & \textbf{2.03} & \textbf{2.51} & 7.41 & 9.23 & \textbf{3.80} & \textbf{4.34}\\
\bottomrule
\end{tabular}%
}
\label{tab:pesq_segsnr_scoreq_static}
\end{table}

\begin{table*}[t]
\caption{Results for the steerable-target scenario. The model name in parentheses indicates the discriminative spatial filtering model used for extracting intermediate features $\Dl$.}
\centering
\small
\begin{tabular}{lllccccccccc}
\toprule
\multicolumn{3}{l}{} 
& \multicolumn{3}{c}{\textbf{PESQ (↑)}} 
& \multicolumn{3}{c}{\textbf{SegSNR (dB) (↑)}} 
& \multicolumn{3}{c}{\textbf{SCOREQ MOS (↑)}} \\
\cmidrule(lr){4-6} \cmidrule(lr){7-9} \cmidrule(lr){10-12}
\textbf{Method} & \textbf{Model} & \textbf{Cond.} 
& -5 dB & 0 dB & 5 dB 
& -5 dB & 0 dB & 5 dB  
& -5 dB & 0 dB & 5 dB \\
\midrule
Unprocessed & Noisy & NA & 1.07 & 1.08 & 1.10 & -12.53 & -8.69 & -5.82 & 1.19 & 1.31 & 1.70 \\
\midrule
Discriminative & JNF \cite{tesch2023multi} & $\Xv,\doa$ & 1.34 & 1.60 & 1.87 & 3.69 & 5.20 & 6.24 & 2.07 & 2.64 & 3.16 \\
Discriminative & DSENet \cite{jing2025end} & $\Xv,\doa$ & 1.56 & 1.88 & 2.19 & 5.00 & \textbf{6.62} & \textbf{7.73} & 2.57 & 3.20& 3.68\\
\midrule
Generative & SpatialGAN & $\Xv,\doa$ & 1.63 & 1.99 & 2.31 & 4.46 & 5.79 & 6.60 & 2.82 & 3.48 & 3.92 \\
\midrule
Gen+Disc & SpatialGAN (JNF) & $\Xv,\Dl$ & 1.71 & 2.06 & 2.35 & 5.04 & 6.29 & 7.05 & 3.28 & 3.90 & 4.21\\
Gen+Disc & SpatialGAN (DSENet) & $\Xv,\Dl$ & 1.68 & 2.05 & 2.35 & 4.81 & 6.17 & 7.11 & 2.94 & 3.61& 4.01 \\
\midrule
Gen+Disc & SpatialGAN (JNF) & $\Xv,\doa,\Dl$ & \textbf{1.78} & \textbf{2.16} & \textbf{2.46} & \textbf{5.19} & 6.49 & 7.28 & \textbf{3.32}& \textbf{3.91} & \textbf{4.24} \\
Gen+Disc & SpatialGAN (DSENet) & $\Xv,\doa,\Dl$ & 1.69 & 2.05 & 2.36 & 4.91 & 6.28& 7.02 & 3.14 & 3.78 & 4.14 \\
\bottomrule
\end{tabular}
\label{tab:steeableresults}
\end{table*}

\section{Experimental Setup}
\label{ED}

\subsection{Dataset Generation}
We generated simulated datasets for training and evaluation with \texttt{pyroomacoustics}~\cite{scheibler2018pyroomacoustics} using the image-source method~\cite{allen1979image}, following~\cite{tesch2023multi}. 
Figure~\ref{fig:simulation_setup} shows the geometric setup. 
Multiple acoustic environments were created with dimensions and reverberation time ($\text{T}_{60}$) uniformly distributed as indicated in Fig.~\ref{fig:simulation_setup}. 
A circular array of three omnidirectional microphones (10~cm diameter) was positioned randomly but at least 1.2~m from the walls in the $x$-$y$ plane at 1.5~m height, and rotated randomly by $\varphi_m \in [0,2\pi)$. The target speaker (red human shape) is placed at \ac{DoA} $\varphi$ relative to the microphone orientation (blue dashed line). 
Interfering speakers were placed in the gray annular region at $0.8$--$1.2$~m distance from the array center, leaving a $10^\circ$ angular gap around the target, and are distributed with one per segment and a minimum $10^\circ$ separation (black human shapes in Fig.~\ref{fig:simulation_setup}).

\subsubsection{Training Dataset}
To simulate the training dataset, we use clean speech from the Interspeech 2020 DNS Challenge dataset~\cite{reddy2020interspeech}. 
We generate two spatial setups: 
(1) \textit{Fixed-target scenario}, where the target speaker's location is always fixed at $\varphi = 0^\circ$, and 
(2) \textit{Steerable-target scenario}, where the target speaker’s position changes across the room. For the fixed-target scenario, we generate approximately $80$ hours of training data. 
In the steerable-target scenario, we fix the \ac{DoA} resolution to $5^\circ$ and simulate around $3.75$ hours of data for each of the $72$ discrete spatial positions, resulting in a total of $\sim270$ hours of training data.

\subsubsection{Test Dataset}

The test samples were $10$~s long. 
For the fixed-target scenario, $600$ samples were generated, yielding $\sim$1.6~hours of data. For the steerable-target scenario, $15$ samples were generated per discrete spatial position, resulting in $\sim$3~hours. 
In both cases, mixture signals were created at three \ac{SNR} levels: $-5$, $0$, and $5$~dB. Interfering speakers and reverberation were treated as noise.

 \subsection{Training Details}
All models were trained at a sampling rate of $16$~kHz. The $\SpatialGAN$ models that were conditioned on intermediate discriminative features $\Dl$ were trained in two steps. 
First, the discriminative spatial filtering models were trained. In our work, we used JNF~\cite{TeschJNF} and DSENet~\cite{jing2025end} as the baselines and conditioning discriminative models. These models were trained on the steerable-target scenario dataset with an FFT length of $512$, window size of $512$, and hop length of $256$. In the original work ~\cite{jing2025end}, DSENet was trained with an FFT length of $256$, window size of $256$, and hop length of $128$; in our setup, we adopted the aforementioned FFT parameters to ensure a fair comparison of the conditioning features in terms of time and frequency resolution, without degrading the original performance.  
Secondly, the $\SpatialGAN$ models were trained using latent features $\Dl$ extracted from the frozen discriminative models, along with the other corresponding conditioning features. For JNF, the intermediate features $\Dl$ were extracted after the F-LSTM block, whereas for DSENet they were extracted after the 4th processing block.  
The \ac{GAN} training followed the procedure described in~\cite{shetu2025leveraging, du2023funcodecfundamentalreproducibleintegrable}, with the following generator STFT parameters: FFT length of $512$, window size of $512$, and hop length of $160$. 

\section{Results}
We adopt JNF~\cite{TeschJNF,tesch2023multi} and DSENet~\cite{jing2025end} as baseline methods from the literature, as mentioned previously. 
The proposed $\SpatialGAN$ is evaluated against these methods using segmental SNR (SegSNR), PESQ~\cite{rix2001perceptual}, and SCOREQ~\cite{ragano2024scoreq} as objective metrics.

As shown in Tab.~\ref{tab:pesq_segsnr_scoreq_static}, in the fixed-target scenario, $\SpatialGAN$ achieves superior PESQ and SCOREQ scores in all SNR groups, improving over DSENet by at least $0.25$ and $0.55$ in the PESQ and SCOREQ metrics, respectively. However, in terms of SegSNR, DSENet shows slightly better metrics by suppressing interferers more aggressively.  

In steerable scenarios, all the variants of $\SpatialGAN$ models also achieve the higher PESQ scores than discrminative baselines. It also outperforms the discriminative baselines in SCOREQ (Tab.~\ref{tab:steeableresults}), indicating a better preservation of speech quality. While DSENet yields a higher SegSNR, this comes at the cost of occasional target speech distortions, evident from the inferior SCOREQ MOS results. Prior studies also suggest that while SegSNR effectively captures energy improvements, it does not always correlate with perceptual quality~\cite{hu2007evaluation,shetu2023ultra}. Our experiments further demonstrate the benefit of incorporating discriminative intermediate features $\Dl$ as conditioning features. As shown in Tab.~\ref{tab:steeableresults}, conditioning $\SpatialGAN$ on both \ac{DoA} $\doa$ and $\Dl$ consistently improves PESQ and SCOREQ by maximum $0.15$ and $0.30$, respectively, compared to conditioning on $\doa$ alone.  We provide demo listening samples here: \url{https://sshetu-iis.github.io/spatialgan/}.

\begin{figure}[t]
\centering
\vspace{-1.5em}
  \input{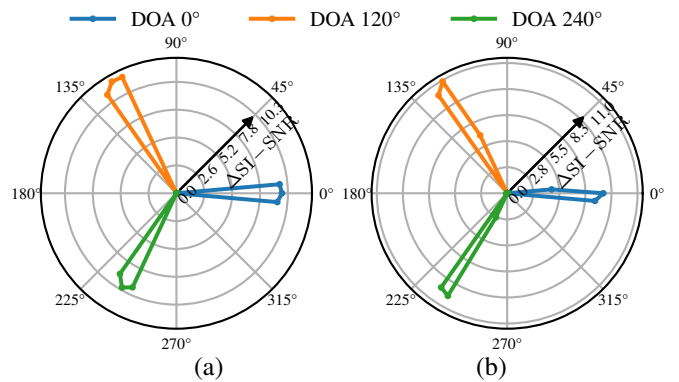}
\vspace{-5em}
\caption{Spatial selectivity in terms of scale-invariant \ac{SNR} (SI-SNR) improvement (arrow toward radial axis) for $\SpatialGAN$ conditioned with (a) ($\Xv,\doa$) and (b) ($\Xv, \doa, \Dl$).}
\label{fig:spatilaselectivity}
\vspace{-1em}
\end{figure}

We additionally analyzed the spatial selectivity of $\SpatialGAN$ (Fig.~\ref{fig:spatilaselectivity}). For this experiment, test samples were selected with target speakers positioned at $\doa_\textrm{target} = 0^\circ$, $120^\circ$, and $240^\circ$ with 5 interferring speakers. The steering \ac{DoA} $\doa_\textrm{steer}$ was swept over $360^\circ$ with $5^\circ$ steps.   The results show that $\SpatialGAN$ can effectively extract speakers in the steered direction while suppressing those from other directions, yieldig positive $\Delta$SI-\ac{SNR}, when $\doa_\textrm{steer}$ matches $\doa_\textrm{target}$.  When conditioned only on $\doa$, the effective beamwidth is approximately $10^\circ$. Including $\Dl$ as additional conditioning further narrows the beamwidth to about $5^\circ$, illustrating sharper spatial selectivity.

\section{Conclusions}
We demonstrate that a GAN can be effectively conditioned with spatial information to extract a target speaker in a steerable manner. The proposed $\SpatialGAN$, conditioned on \ac{DoA} and intermediate discriminative feature representations, consistently outperforms discriminative baseline methods in speech quality metrics while maintaining strong spatial selectivity. 


\section{Acknowledgements}
The authors gratefully acknowledge the scientific support and HPC resources provided by the Erlangen National High Performance Computing Center (NHR@FAU) of the Friedrich-Alexander-Universität Erlangen-Nürnberg (FAU). The hardware is funded by the German Research Foundation (DFG).

\let\oldbibliography\thebibliography
\renewcommand{\thebibliography}[1]{%
  \oldbibliography{#1}%
  \footnotesize 
  \setlength{\itemsep}{-0.0ex} 
  \setlength{\parsep}{0pt}  
  \setlength{\parskip}{0pt} 
  \setlength{\leftmargin}{1em} 
}

\bibliographystyle{IEEEbib}
\bibliography{strings,refs}

\begin{thebibliography}{10}

\bibitem{cherry1953some}
E.~C. Cherry,
\newblock ``Some experiments on the recognition of speech, with one and with two ears,''
\newblock {\em J. Acoust. Soc. Am.}, 1953.

\bibitem{pollack1957cocktail}
I.~Pollack and J.~M. Pickett,
\newblock ``Cocktail party effect,''
\newblock {\em J. Acoust. Soc. Am.}, 1957.

\bibitem{kovalyov2023dsenet}
A.~Kovalyov, K.~Patel, and I.~Panahi,
\newblock ``{DSENet}: Directional signal extraction network for hearing improvement on edge devices,''
\newblock {\em IEEE Access}, 2023.

\bibitem{10694623}
M.~Strauss and O.~Köpüklü,
\newblock ``Efficient area-based and speaker-agnostic source separation,''
\newblock in {\em Proc. Int. Workshop Acoust. Signal Enhanc.}, 2024.

\bibitem{li2014overview}
J.~Li, Li~Deng, Y.~Gong, and R.~Haeb-Umbach,
\newblock ``An overview of noise-robust automatic speech recognition,''
\newblock {\em IEEE/ACM Trans. Audio, Speech, Language Process.}, 2014.

\bibitem{5109760}
S.~Markovich, S.~Gannot, and I.~Cohen,
\newblock ``Multichannel eigenspace beamforming in a reverberant noisy environment with multiple interfering speech signals,''
\newblock {\em IEEE/ACM Trans. Audio, Speech, Language Process.}, 2009.

\bibitem{gannot2017consolidated}
S.~Gannot, E.~Vincent, S.~Markovich-Golan, and A.~Ozerov,
\newblock ``A consolidated perspective on multimicrophone speech enhancement and source separation,''
\newblock {\em IEEE/ACM Trans. Audio, Speech, Language Process.}, 2017.

\bibitem{5067369}
E.~A.~P. Habets, J.~Benesty, I.~Cohen, S.~Gannot, and J.~Dmochowski,
\newblock ``New insights into the {MVDR} beamformer in room acoustics,''
\newblock {\em IEEE/ACM Trans. Audio, Speech, Language Process.}, 2010.

\bibitem{gu2019neural}
R.~Gu, L.~Chen, S.-X. Zhang, J.~Zheng, Y.~Xu, M.~Yu, D.~Su, Y.~Zou, and D.~Yu,
\newblock ``Neural spatial filter: Target speaker speech separation assisted with directional information.,''
\newblock in {\em Proc. INTERSPEECH}, 2019.

\bibitem{benesty2008microphone}
J.~Benesty, J.~Chen, and Y.~Huang,
\newblock {\em Microphone array signal processing},
\newblock Springer, 2008.

\bibitem{ephrat2018looking}
A.~Ephrat, I.~Mosseri, O.~Lang, K.~Dekel, T.and~Wilson, A.~Hassidim, W.~T Freeman, and M.~Rubinstein,
\newblock ``Looking to listen at the cocktail party: a speaker-independent audio-visual model for speech separation,''
\newblock {\em ACM Trans. Graph.}, 2018.

\bibitem{alfouras2018conversation}
T.~Alfouras, J.S. Chung, and A.~Zisserman,
\newblock ``The conversation: deep audio-visual speech enhancement,''
\newblock in {\em Proc. INTERSPEECH}, 2018.

\bibitem{shetu2021empirical}
S.~S. Shetu, S.~Chakrabarty, and E.~A.~P Habets,
\newblock ``An empirical study of visual features for {DNN} based audio-visual speech enhancement in multi-talker environments,''
\newblock in {\em Proc. IEEE Int. Conf. Acoust., Speech, Signal Process.}, 2021.

\bibitem{zmolikova2023neural}
K.~Zmolikova, M.~Delcroix, T.~Ochiai, K.~Kinoshita, J.~{\v{C}}ernock{\`y}, and D.~Yu,
\newblock ``Neural target speech extraction: An overview,''
\newblock {\em IEEE Signal Process. Mag.}, 2023.

\bibitem{elminshawi2024new}
M.~Elminshawi, W.~Mack, S.~R. Chetupalli, S.~Chakrabarty, and E.~A.~P Habets,
\newblock ``New insights on the role of auxiliary information in target speaker extraction,''
\newblock {\em Front. Signal Process.}, 2024.

\bibitem{ji2020speaker}
X.~Ji, M.~Yu, C.~Zhang, D.~Su, T.~Yu, X.~Liu, and D.~Yu,
\newblock ``Speaker-aware target speaker enhancement by jointly learning with speaker embedding extraction,''
\newblock in {\em Proc. IEEE Int. Conf. Acoust., Speech, Signal Process.}, 2020.

\bibitem{elminshawi2023beamformer}
M.~Elminshawi, S.~R. Chetupalli, and E.~A.~P. Habets,
\newblock ``Beamformer-guided target speaker extraction,''
\newblock in {\em Proc. IEEE Int. Conf. Acoust., Speech, Signal Process.}, 2023.

\bibitem{TeschJNF}
K.~Tesch and T.~Gerkmann,
\newblock ``Spatially selective deep non-linear filters for speaker extraction,''
\newblock in {\em Proc. IEEE Int. Conf. Acoust., Speech, Signal Process.}, 2023.

\bibitem{tesch2023multi}
K.~Tesch and T.~Gerkmann,
\newblock ``Multi-channel speech separation using spatially selective deep non-linear filters,''
\newblock {\em IEEE/ACM Trans. Audio, Speech, Language Process.}, 2023.

\bibitem{jing2025end}
K.~Jing, W.~Zhang, and Y.~Gao,
\newblock ``End-to-end {DOA}-guided speech extraction in noisy multi-talker scenarios,''
\newblock {\em Proc. INTERSPEECH}, 2025.

\bibitem{jebara2012machine}
T.~Jebara,
\newblock {\em Machine learning: discriminative and generative},
\newblock Springer, 2012.

\bibitem{shetu2024comparative}
S.~S. Shetu, E.~A.~P. Habets, and A.~Brendel,
\newblock ``Comparative analysis of discriminative deep learning-based noise reduction methods in low {SNR} scenarios,''
\newblock in {\em Proc. Int. Workshop Acoust. Signal Enhanc.}, 2024, pp. 36--40.

\bibitem{fu2019metricgan}
S.-W. Fu, C.-F. Liao, Y.~Tsao, and S.-D. Lin,
\newblock ``{MetricGAN}: Generative adversarial networks based black-box metric scores optimization for speech enhancement,''
\newblock in {\em Proc. Int. Conf. Mach. Learn. (ICML)}, 2019.

\bibitem{sani2023improving}
P.~Sani, J.~Bauer, F.~Zalkow, E.~A.~P. Habets, and C.~Dittmar,
\newblock ``Improving the naturalness of synthesized spectrograms for {TTS} using {GAN}-based post-processing,''
\newblock in {\em Proc. ITG Speech Commun. Conf.}, 2023.

\bibitem{shetu2025gan}
S.~S. Shetu, E.~A.~P. Habets, and A.~Brendel,
\newblock ``{GAN}-based speech enhancement for low {SNR} using latent feature conditioning,''
\newblock in {\em Proc. IEEE Int. Conf. Acoust., Speech, Signal Process.}, 2025.

\bibitem{chen2025tfdense}
H.~Chen, J.~Zhang, Y.~Fu, X.~Zhou, R.~Wang, Y.~Xu, and D.~Ke,
\newblock ``{TFDense-GAN}: a generative adversarial network for single-channel speech enhancement,''
\newblock {\em EURASIP J. Adv. Signal Process.}, 2025.

\bibitem{yu2024ks}
G.~Yu, R.~Han, C.~Xu, H.~Zhao, N.~Li, C.~Zhang, X.~Zheng, C.~Zhou, Q.~Huang, and B.~Yu,
\newblock ``{KS-Net}: Multi-band joint speech restoration and enhancement network,''
\newblock in {\em Proc. IEEE Int. Conf. Acoust., Speech, Signal Process.}, 2024.

\bibitem{10446418}
L.~Yu, W.~Zhang, C.~Du, L.~Zhang, Zheng L., and Y.~Qian,
\newblock ``Generation-based target speech extraction with speech discretization and vocoder,''
\newblock in {\em Proc. IEEE Int. Conf. Acoust., Speech, Signal Process.}, 2024.

\bibitem{kamo2023target}
M.~Kamo, N.and~Delcroix and T.~Nakatani,
\newblock ``Target speech extraction with conditional diffusion model,''
\newblock {\em Proc. INTERSPEECH}, 2023.

\bibitem{ma2025enhancing}
H.~Ma, R.~Chen, X.~Zhang, J.~Liu, and X.~Li,
\newblock ``Enhancing intelligibility for generative target speech extraction via joint optimization with target speaker {ASR},''
\newblock {\em IEEE Signal Process. Lett.}, 2025.

\bibitem{navon2025flowtse}
A.~Navon, A.~Shamsian, Y.~Segal-Feldman, N.~Glazer, G.~Hetz, and J.~Keshet,
\newblock ``{FlowTSE}: Target speaker extraction with flow matching,''
\newblock {\em arXiv preprint}, 2025.

\bibitem{shetu2025leveraging}
S.~S. Shetu, E.~A.~P. Habets, and A.~Brendel,
\newblock ``Leveraging discriminative latent representations for conditioning {GAN}-based speech enhancement,''
\newblock {\em arXiv preprint}, 2025.

\bibitem{du2023funcodecfundamentalreproducibleintegrable}
Z.~Du, S.~Zhang, K.~Hu, and S.~Zheng,
\newblock ``{FunCodec}: A fundamental, reproducible and integrable open-source toolkit for neural speech codec,''
\newblock in {\em Proc. IEEE Int. Conf. Acoust., Speech, Signal Process.}, 2023.

\bibitem{tagliasacchi2020seanetmultimodalspeechenhancement}
M.~Tagliasacchi, Y.~Li, K.~Misiunas, and D.~Roblek,
\newblock ``{SEANet}: A multi-modal speech enhancement network,''
\newblock in {\em Proc. INTERSPEECH}, 2020.

\bibitem{defossez2022high}
A.~Défosséz, J.~Copet, G.~Synnaeve, and Y.~Adi,
\newblock ``High fidelity neural audio compression,''
\newblock {\em Trans. Mach. Learn. Res.}, 2023.

\bibitem{perez2018film}
E.~Perez, F.~Strub, H.~de~Vries, V.~Dumoulin, and A.~Courville,
\newblock ``{FiLM}: Visual reasoning with a general conditioning layer,''
\newblock in {\em Proc. AAAI Conf. Artif. Intell.}, 2018.

\bibitem{scheibler2018pyroomacoustics}
R.~Scheibler, E.~Bezzam, and I.~Dokmani{\'c},
\newblock ``Pyroomacoustics: A python package for audio room simulation and array processing algorithms,''
\newblock in {\em Proc. IEEE Int. Conf. Acoust., Speech, Signal Process.}, 2018.

\bibitem{allen1979image}
J.~B. Allen and D.~A. Berkley,
\newblock ``Image method for efficiently simulating small-room acoustics,''
\newblock {\em J. Acoust. Soc. Am.}, 1979.

\bibitem{reddy2020interspeech}
C.~K.~A. Reddy, V.~Gopal, R.~Cutler, E.~Beyrami, R.~Cheng, H.~Dubey, S.~Matusevych, R.~Aichner, A.~Aazami, S.~Braun, et~al.,
\newblock ``The {InterSpeech} 2020 deep noise suppression challenge: Datasets, subjective testing framework, and challenge results,''
\newblock in {\em Proc. INTERSPEECH}, 2020.

\bibitem{rix2001perceptual}
A.~W. Rix, J.~G. Beerends, M.~P. Hollier, and A.~P. Hekstra,
\newblock ``Perceptual evaluation of speech quality ({PESQ})-a new method for speech quality assessment of telephone networks and codecs,''
\newblock in {\em Proc. IEEE Int. Conf. Acoust., Speech, Signal Process.}, 2001.

\bibitem{ragano2024scoreq}
A.~Ragano, J.~Skoglund, and A.~Hines,
\newblock ``{SCOREQ}: Speech quality assessment with contrastive regression,''
\newblock {\em Proc. Adv. Neural Inf. Process. Syst.}, 2024.

\bibitem{hu2007evaluation}
Y.~Hu and P.~C. Loizou,
\newblock ``Evaluation of objective quality measures for speech enhancement,''
\newblock {\em IEEE/ACM Trans. Audio, Speech, Language Process.}, 2007.

\bibitem{shetu2023ultra}
S.~S. Shetu, S.~Chakrabarty, O.~Thiergart, and E.~Mabande,
\newblock ``Ultra low complexity deep learning based noise suppression,''
\newblock in {\em Proc. IEEE Int. Conf. Acoust., Speech, Signal Process.}, 2024.

\end{thebibliography}

\end{document}